This article has been accepted for publication in IEEE Transactions on Energy Conversion. This is the author's version which has not been fully edited and content may change prior to final publication. Citation information: DOI 10.1109/TEC.2025.3556100

# Model-Agnostic Meta-Learning for Fault Diagnosis of Induction Motors in Data-Scarce Environments with Varying Operating Conditions and Electric Drive Noise

Ali Pourghoraba, MohammadSadegh KhajueeZadeh, Ali Amini, Abolfazl Vahedi, *Senior Member*, *IEEE*, Gholam Reza Agah, and Akbar Rahideh

***Abstract*— Reliable mechanical fault detection with limited data is crucial for the effective operation of induction machines, particularly given the real-world challenges present in industrial datasets, such as significant imbalances between healthy and faulty samples and the scarcity of data representing faulty conditions. This research introduces an innovative meta-learning approach to address these issues, focusing on mechanical fault detection in induction motors across diverse operating conditions while mitigating the adverse effects of drive noise in scenarios with limited data. The process of identifying faults under varying operating conditions is framed as a few-shot classification challenge and approached through a model-agnostic meta-learning strategy. Specifically, this approach begins with training a meta-learner across multiple interconnected fault-diagnosis tasks conducted under different operating conditions. In this stage, cross-entropy is utilized to optimize parameters and develop a robust representation of the tasks. Subsequently, the parameters of the meta-learner are fine-tuned for new tasks, enabling rapid adaptation using only a small number of samples. This method achieves excellent accuracy in fault detection across various conditions, even when data availability is restricted. The findings indicate that the proposed model outperforms other sophisticated techniques, providing enhanced generalization and quicker adaptation. The accuracy of fault diagnosis reaches a minimum of 99%, underscoring the model's effectiveness for reliable fault identification.**

***Index Terms*—Mechanical fault detection, induction motors, meta-learning, noisy environments, limited data, varying operating conditions.**

## I. INTRODUCTION

INDUCTION motors are widely used across various industrial sectors due to their straightforward configuration, ease of installation, and cost-effective maintenance [1]. However, failures can occur after prolonged operation, potentially leading to malfunctions. Components such as bearings, stator windings, and rotor bars are particularly vulnerable to faults, which can result in significant economic losses and even pose safety risks. As a result, ensuring the safety and stability of induction motors has become increasingly important, making early fault diagnosis essential for preventing disasters [2].

Currently, vibration signals are a key focus for monitoring the condition of rotating machinery because they provide critical insights into operating conditions. These signals are vital for assessing performance and identifying potential issues, making them indispensable for effective maintenance strategies [3]. However, the need for expensive sensors to be installed on bearing housings, along with the requirement for direct access to machinery for measurements, presents significant drawbacks. These limitations restrict the use of vibration signals in systems that are not easily accessible [1].

In the industrial sector, many devices that utilize induction motors already incorporate current monitoring for control functions through frequency inverters or for protection purposes with the help of current transformers. As a result, the motor's stator current is typically available for fault diagnosis, eliminating the need for additional sensor installations. This accessibility simplifies the process of identifying issues and enhances maintenance efficiency [4].

In recent decades, machine learning has become ubiquitous in the fault diagnosis of induction motors. These methods can reduce computational memory requirements while maintaining high diagnostic accuracy. However, effective feature extraction is critical for determining model performance, making it an essential component of the fault diagnosis process [5]. Additionally, as technology continues to advance, vast quantities of data from various operating conditions have been gathered. Traditional approaches often struggle to analyze these large datasets due to the extensive processing time required [1], [2]. While the aforementioned techniques can be useful for detecting faults in induction motors, their effectiveness and quality heavily depend on prior experience and are typically suited for specific tasks. Therefore, there is a pressing need to investigate more objective and efficient methods [6].

In recent years, deep neural networks have significantly advanced their ability to extract fault features and accurately

Ali Pourghoraba, MohammadSadegh KhajueeZadeh, Ali Amini, and Abolfazl Vahedi are with the Department of Electrical Engineering, Iran University of Science and Technology, Tehran 16846-13114, Iran (e-mail: ali_pour98@elec.iust.ac.ir; mohammadkhajuee@yahoo.com; al_amini76@elec.iust.ac.ir; avahevi@iust.ac.ir;).

Gholam Reza Agah and Akbar Rahideh are with the Department of Electrical Engineering, Shiraz University of Technology, Shiraz 13876-71557, Iran (e-mail: reza.agah.8742@gmail.com; rahide@sutech.ac.ir).

predict fault categories using extensive labeled datasets [7], [8]. However, gathering a substantial number of labeled samples in real-world industrial scenarios remains challenging. Models with numerous parameters often overfit when trained on limited sample sizes, while those with insufficient parameters may lack the diagnostic proficiency required. Conversely, models with too few parameters demonstrate inadequate diagnostic effectiveness [9]. Consequently, achieving accurate fault diagnosis with a minimal set of labeled samples presents a considerable hurdle [10].

To address the challenges associated with limited datasets, one effective strategy is data augmentation. Yong Oh et al. [11] introduced a method utilizing generative adversarial networks (GANs) capable of producing high-quality synthetic data that closely resembles real-world samples. This approach is particularly noteworthy due to the powerful data generation capabilities inherent in GAN architectures. The results demonstrate that this method successfully mitigates issues related to data imbalance and the limitations of having few labeled samples [12]. While GANs offer significant benefits in addressing these challenges, the quality of the synthetic data they produce warrants further examination. Additionally, this approach may struggle with generalization when faced with changing operating conditions [13].

To tackle this issue, transfer learning has been employed, allowing for the adaptation of knowledge from a source domain to enhance model performance. By incorporating additional examples from the target domain, this approach effectively addresses differences in data distributions [14]. Transfer learning encompasses a set of machine learning techniques that leverage existing models or datasets to improve performance on new, related tasks. This strategy enables practitioners to build upon prior knowledge, facilitating more efficient learning and adaptation in various contexts [15].

For transfer learning to be effective in fault diagnosis, it is essential that the data originates from the same or closely related mechanical components. Consequently, both the training and testing datasets must encompass the same fault categories to ensure compatibility and accuracy in the diagnostic process [16]. In practical industrial scenarios, it is common to have a significant amount of labeled fault data available from a single component for training models. However, challenges arise when attempting to apply this model to detect fault types in another component that has considerably fewer labeled instances. The discrepancies in fault categories between the source and target components limit the effectiveness of traditional transfer learning methods in addressing this issue [17].

To tackle the challenges associated with limited data and improve the generalization of fault diagnosis models under complex operating conditions, this research advocates for a model-agnostic meta-learning strategy specifically designed for detecting faults in electric machines [18]. The model-agnostic nature of meta-learning allows it to adapt to various learning scenarios reliant on gradient-based optimization. By developing a foundational learner, meta-learning aims to facilitate rapid adaptation to new tasks with minimal exposure to unseen data [19]. A meta-learner guides this foundational learner, equipping it with the skills necessary to adjust swiftly to new challenges [19], [20]. Proper training of initial parameters is crucial, as it enables effective gradient updates from limited task data [21]. This research applies model-agnostic meta-learning to fault detection in squirrel-cage induction motors, treating varying operating conditions as distinct multiclass classification tasks [22], [23].

Altae-Tran et al. [24] utilized one-shot learning techniques for drug discovery in scenarios with limited data availability. Snell et al. [25] introduced prototypical networks as a method for few-shot learning, establishing a metric space that facilitates classification by measuring distances to class prototypes. This approach enables effective identification of classes based on limited examples. Qiao et al. [26] developed a few-shot image recognition method that predicts parameters based on activation patterns. Zhang et al. [27] proposed a straightforward yet versatile framework known as MetaGAN to address few-shot learning challenges. Unlike traditional semi-supervised few-shot learning methods, their algorithms can handle semi-supervised tasks at both the individual sample and overall task levels. Furthermore, as a meta-learning approach, this method effectively mitigates the overfitting issue that arises from having a small sample size. Consequently, the model-agnostic meta-learning technique employed in this research presents a promising strategy for diagnosing faults in electric machines [19]-[23].

The remainder of this study is organized as follows: Section II provides an in-depth overview of the underlying theory and framework of the proposed model. Section III outlines the experimental procedures and includes a comprehensive analysis of the results. Finally, Section IV presents a summary of the key findings and conclusions drawn from the research.

## II. Basic Method And Proposed Model

### A. Basic Theory of Model-agnostic Meta-learning

In few-shot learning, the data is organized into two main components: the meta-training set (also known as the support set), denoted as $S = \{(I_n^{train}.I_n^{test})\}_{n=1}^{N}$, and the meta-testing set (or query set), represented as $Q = \{(I_m^{train}.I_m^{test})\}_{m=1}^{M}$, where N and M are the sizes of the support set and query set, respectively. The key characteristic of these datasets is that they consist of entirely different categories. Each task in the meta-testing set, represented by $I_m^{train}$, contains a small number of images. It should be noted that the current signals used in this work are represented as $I$ . The training images are denoted as $I^{train} = \{(x_j.y_j)\}_{j=1}^{J}$ and the test images as $I^{test} = \{(x_p.y_p)\}_{p=1}^{P}$, where $x$ represents the input images and $y$ represents their corresponding labels. Additionally, J and P are the sizes of the training and test data in each set. Importantly, both the training and testing images are drawn from the same underlying distribution, which allows for effective generalization despite the limited number of training examples.

A base learner $\mathcal{A}$ is defined by the function $y_* = f_\theta(x_*)$, where $y_*$ represents the predicted output for a given input $x_*$, and $\theta$ denotes the parameters of the model. This base learner is trained using a training dataset and evaluated on a separate testing dataset. To enhance the representation of the input data $x_*$, both the training and testing samples are transformed into a feature space using an embedding function defined as $\varphi_* = f_\varphi(x_*)$. The primary objective of few-shot learning methods is to develop an effective embedding function that minimizes the average test error of the base learner across a diverse range of tasks.

Few-shot learning addresses scenarios where only a small number of labeled samples are available for training. By leveraging a well-designed embedding function $f_\varphi$, these methods generalize effectively across different tasks, even with limited data. The embedding function $f_\varphi$ , parameterized by $\varphi$, is a learnable model that maps the input data $x_*$ into a feature space where the base learner $f_\theta$ can operate more effectively. Specifically, $\varphi_* = f_\varphi(x_*)$ represents the transformed feature embedding of the input $x_*$. Here, $\varphi_*$ is the output of the embedding function and serves as a compact, task-relevant representation of the input.

The parameter $\theta$ governs the base learner, which makes predictions based on the embeddings generated by $f_\varphi$. While $\varphi$ focuses on creating a meaningful and generalizable representation of the input across tasks, $\theta$ adapts to the specific task using the embeddings. Thus, $f_\varphi$ provides the foundation for generalization across tasks, while $f_\theta$ focuses on task-specific inference. The primary goal of few-shot learning is to learn the parameters $\varphi$ such that the embedding function $f_\varphi$ minimizes the average test error of the base learner $f_\theta$ across a range of tasks. In this way, $\varphi$ captures universal features that are transferable, enabling the base learner to perform well even with limited task-specific data. This interplay between $\varphi$, $\varphi_*$ , and $\theta$ is central to the effectiveness of few-shot learning systems. This objective can be expressed as follows:

$$\varphi = argmin_\varphi \mathrm{E}_T[L^{meta}(I^{test}; \theta, \varphi)] \qquad (1)$$

Here, $\theta = \mathrm{A}(I^{train}. \varphi)$, the $argmin_\varphi$ operation finds the value of $\varphi$ that minimizes the expression inside the brackets. The $\mathrm{E}_T[.]$ denotes the expected value over the task distribution $T$, and $L^{meta}$ is the meta-loss function that evaluates the performance of $f_\theta$ on the test dataset $I^{test}$. The performance of the model across tasks is assessed by computing:

$$\theta(\varphi) = E_S[L^{meta}(I^{test}; \theta, \varphi)] \qquad (2)$$

where $E_S[.]$ represents the expected value over tasks sampled from the meta-training dataset $S$. The embedding function $f_\varphi$ is further refined during training by minimizing a classification loss, as defined by:

$$\varphi = argmin_\varphi L^{ce}(I^{test}; \varphi) \qquad (3)$$

The training process begins with the optimization of a cross-entropy loss function, denoted as $L^{ce}$. This ensures that the learned embeddings are both discriminative and generalizable, enabling the base learner $f_\theta$ to perform well on unseen tasks. Together, the embedding function $f_\varphi$ and the task-specific learner $f_\theta$ work in harmony: $f_\varphi$ provides transferable features, while $f_\theta$ specializes in task-specific predictions, achieving robust generalization in few-shot learning settings. Once the initial embedding model $\varphi$ is trained, it is refined further by developing a subsequent embedding model $\varphi'$ to enhance the feature space.

This refinement is achieved by minimizing a combined objective function that balances two components: the cross-entropy loss $L^{ce}$ and the Kullback-Leibler (KL) divergence, $\nabla_{KL}$. The cross-entropy loss measures the discrepancy between the predicted outputs $f(I^{train}; \varphi')$ from the new embedding model and the ground-truth labels in the dataset $I^{train}$. This ensures accurate alignment of predictions with the true class labels. The KL divergence, on the other hand, quantifies how closely the predicted output distribution $f(I^{train}; \varphi')$ matches a "soft target" distribution derived from the initial embedding model $f(I^{train}; \varphi)$. These soft targets capture additional information about inter-class relationships, helping to regularize the learning process and encourage smoother, more meaningful feature representations. The combined objective function for the new embedding model is expressed as (4) where the $\alpha$ and $\beta$ are hyperparameters that balance the importance of each loss component.

This phase of the model development resembles a self-distillation process, where the teacher and student networks share the same architecture. The refined embedding model $f_{\varphi'}$, developed in earlier training, serves as a fixed feature extractor. After completing this phase, the model's generalization ability is evaluated using a meta-testing dataset. Following this, the foundational learner, parameterized by $\theta = \{W. b\}$ , is trained on a specific task $I^{train}$, which is drawn from the meta-testing distribution. The optimization objective for the foundational learner is given by (5), where $W$ is the weight matrix, $b$ is the bias vector, $L_t^{ce}$ is the cross-entropy loss measuring the discrepancy between predictions $Wf_{\varphi'}(x_t) + b$ and true labels $y_t$, and $R(W. b)$ is the $L_2$ regularization term that penalizes large values of $W$ and $b$ to prevent overfitting. The fixed embedding model $f_{\varphi'}(x_t)$ transforms input $x_t$ into robust feature representations. This process ensures the foundational learner adapts effectively to the task-specific data $I^{train}$, leveraging the embedding model to provide generalized, transferable features. Regularization through $R(W. b)$ enhances stability and smoothness in the learning process.

Expanding this approach, each fault detection scenario under varying operating conditions is treated as a distinct task within the meta-learning framework. The model demonstrates exceptional versatility, enabling it to generalize across diverse fault detection scenarios instead of being confined to a single application. This adaptability allows the model to achieve high performance even when faced with a new fault detection challenge, requiring only a small dataset and minor adjustments through gradient-based learning techniques. Such flexibility underscores the meta-learning framework's capacity to address real-world applications where rapid and efficient task

$$\varphi' = argmin_{\varphi'}(\alpha L^{ce}(I^{train}; \varphi') + \beta \nabla_{KL}(f(I^{train}; \varphi'), f(I^{train}; \varphi'))) \qquad (4)$$

$$\theta = argmin_{\{W.b\}} \sum_{t=1}^{T} L_t^{ce}\left(Wf_{\varphi'}(x_t) + b. y_t\right) + R(W. b) \qquad (5)$$

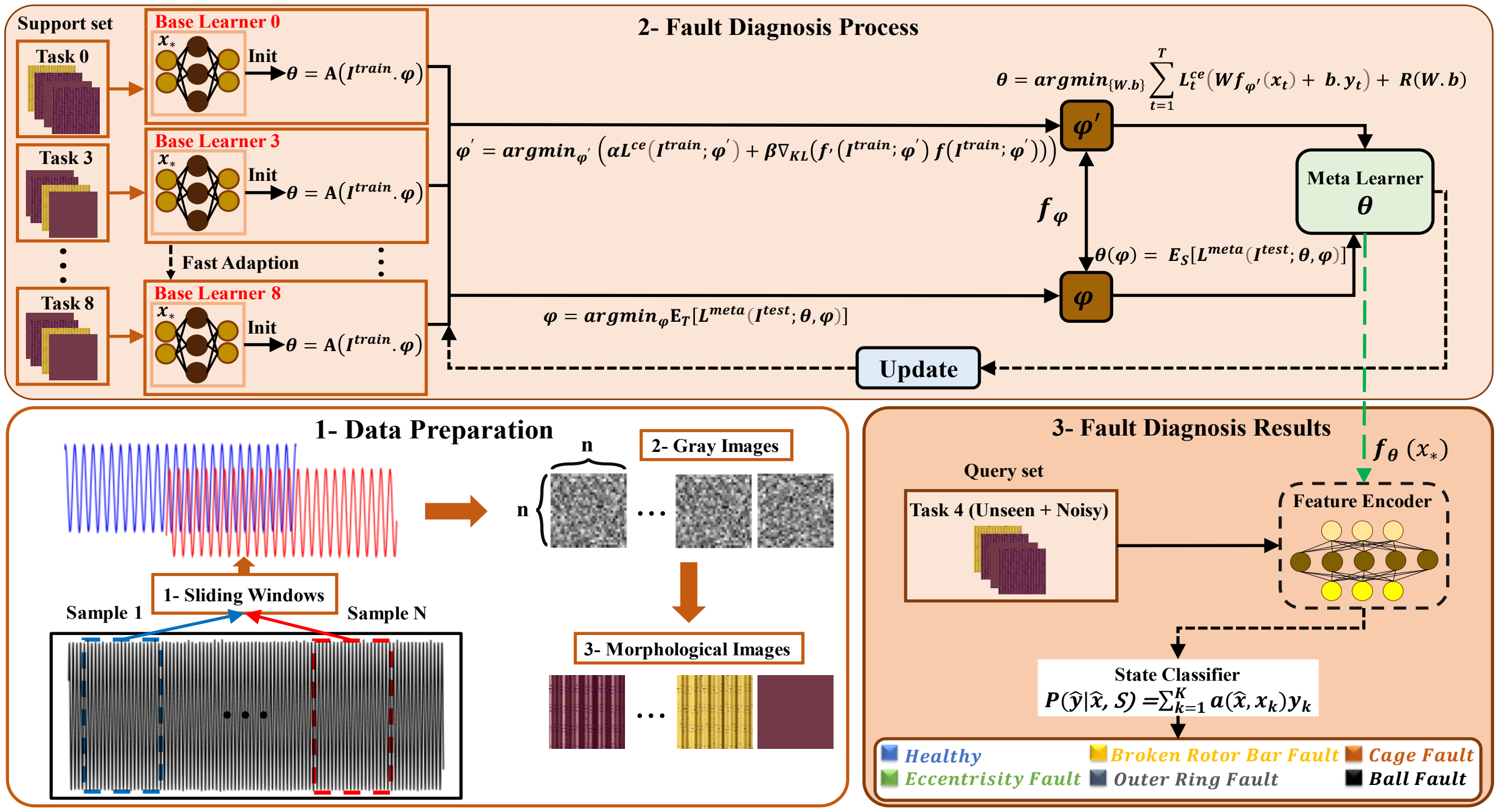

Fig. 1. Proposed methodological framework for intelligent fault diagnosis.

adaptation is essential. Further details about the model's architecture and its practical applications will be explored in subsequent sections.

Furthermore, (6) represents a key aspect of few-shot learning, specifically in the context of metric-based approaches. This equation calculates the probability of a predicted label $\hat{y}$ for a query point $\hat{x}$, given a support set $S$. Here, $a(\hat{x}, x_k)$ is an attention mechanism or similarity function that measures the relevance of each support example $x_k$ to the query, and $y_k$ is the corresponding label. The summation over $K$ support examples allow the model to make predictions based on the weighted influence of similar instances in the support set, embodying the essence of learning from few examples in few-shot learning.

$$P(\hat{y}|\hat{x}, S) = \sum_{k=1}^{K} a(\hat{x}, x_k) y_k \qquad (6)$$

*B. Structure and Training Methodology of the Suggested Model*

The overall architecture and process of the proposed model-agnostic meta-learning-based model are illustrated in Fig. 1 and Algorithm 1, respectively. The process begins with the collection of motor current signals, where single-phase current signals are used for mechanical fault diagnosis instead of conventional three-phase signals due to their practicality, lower hardware complexity, and computational efficiency. Using single-phase signals eliminates the need for additional sensors, thereby reducing installation and maintenance costs. It also simplifies data processing by avoiding redundancy in three-phase signals, which leads to lower storage and computational requirements. This approach is particularly beneficial for real-time applications and edge computing, enabling efficient fault detection while maintaining high diagnostic accuracy. Then, a sliding window technique is employed to divide the signal into uniform samples. The input current signals are initially one-dimensional. Since our model requires a 2D image representation of the signals, these signals are transformed into a 2D square matrix by reorganizing the arrays in the signal domain. Assuming each original signal is represented as an $1 \times n^2$ array, the transformation results in a $n \times n$ matrix. This conversion allows the model to efficiently process the signals in a 2D format, enhancing its ability to capture spatial features and improve overall performance in the analysis, as shown in (7).

$$[I_1 \quad \cdots \quad I_{n^2}] \rightarrow \begin{bmatrix} I_1 & \cdots & I_n \\ \vdots & \ddots & \vdots \\ I_{(n-1)(n+2)} & \cdots & I_{n^2} \end{bmatrix} \qquad (7)$$

In our recommended methodology, each fragment of the current signal, with a length of $1 \times n^2$, is reorganized into a 2D square matrix of $n \times n$. This transformation effectively creates a grayscale image that represents the corresponding current signal. After that, we will use the morphological method to enhance efficiency by revealing details in the images.

Morphological image processing encompasses a collection of nonlinear techniques designed to analyze and manipulate the shapes and structures within an image [28]. These methods focus on the arrangement of pixel values rather than their numerical quantities, making them particularly effective for various image processing tasks. They can be applied to grayscale images even when light transfer functions are unknown, as the actual pixel values may often hold minimal significance. Morphological methods employ a small shape or template known as a structuring element [29], which is systematically positioned at different locations throughout the image to facilitate comparisons with surrounding pixel neighborhoods [30]. Certain operations determine whether the structuring element is contained within these neighborhoods or overlaps with them, thereby enhancing the precision of our

**Algorithm 1** MAML for few-shot Supervised Learning

**Input:** Support set $S = \{(I_n^{train}.I_n^{test})\}_{n=1}^{N}$ and query set $Q = \{(I_m^{train}.I_m^{test})\}_{m=1}^{M}$ and weights and bias vector $(W, b)$ and base learner $\mathcal{A}$.

**Output:** Predicted health state of the motor.

**Require:** $\theta(\tau)$: distribution over tasks (which $\tau$ stands for task)
**Require:** $\alpha, \beta$: step size hyperparameters
randomly initialized $\theta$
**while** not done **do**
    Sampled batch of tasks $\tau_i \sim \theta(\tau)$
    **for** all $\tau_i$ **do**
        Sample $J$ data points $I = (x_j.y_j)$ from $\tau_i$
        Evaluate $argmin_\varphi E_T[L^{meta}(I)]$ using $\theta$ and $\varphi$
        $L_{\tau_i} \leftarrow loss_{task}(\theta(\varphi_*.x^{(i)}{}_j).y^{(i)}{}_j)$
        Compute adapted parameters with minimum cross-entropy: $\varphi_i = argmin_{\varphi_i} L^{ce}(I^{test}; \varphi_i)$
        Sample data points $I = (x_p.y_p)$ from $\tau_i$ for the meta-update
    **end for**
        Update $\theta \leftarrow \theta - argmin_{\{W.b\}} \Sigma_{t=1}^{T} L_t^{ce}\left(Wf_{\varphi'}(x_t) + b.y_t\right) + R(W.b)$ using each $I_i^{test}$ and $L_{\tau_i}$
    $L_{\tau_i} \leftarrow L_{\tau_i} + loss_{task}(\theta(\varphi_*.x^{(i)}{}_j).y^{(i)}{}_j)$
**end while**

networks [31], as discussed in the experiments and analysis of results. This approach offers distinct advantages over wavelet transforms and Fast Fourier Transform (FFT), particularly in the analysis of shape and structure. Operations such as erosion and dilation are robust against noise, enabling effective object segmentation and boundary extraction while preserving essential features. Furthermore, these techniques are intuitive and computationally efficient, making them well-suited for local structural analysis in both binary and grayscale images, as well as for examining complex shapes and textures.

## III. Experiments And Analysis of Results

### *A. Setup and Measurements Procedures*

Broken bars, bearing faults, and eccentricity are among the most well-known fault conditions categorized within mechanical groups that can occur in induction motors. Early identification of these faults is crucial to prevent further damage to the motor and minimize costly downtime. Approximately 80% of mechanical faults lead to eccentricity [32]. It is important to note that eccentricity can arise from manufacturing and assembly processes due to various factors, including uneven air gaps between the stator core and rotor, manufacturing defects, rotor misalignment, and bearing wear [19], [32].

Conversely, mechanical stresses can be caused by overloads, sudden load changes, high temperatures, and excessive mechanical loading—particularly during startup—as well as defective casting or weak joints during the manufacturing process, which can lead to broken rotor bars [33]. The causes of bearing failures are similar; up to 80% of these failures are attributed to improper lubrication, damage from vibration, issues stemming from incorrect installation, excessive heat, high mechanical loading (especially during startup), and manufacturing defects such as poor casting or weak joints [34].

The test bench designed for mechanical fault detection, as depicted in Fig. 2, consists of several critical components carefully configured to effectively simulate diverse fault conditions in induction motors. At the core of this setup is a 1.5 kW squirrel-cage induction motor coupled mechanically to a 3 kW self-excited synchronous generator. Details of the tested induction motor are provided in Table I. The driving-end ball bearing in the investigated induction motor is of the 6205 NTN category, with specific data outlined in Table II. The generator output is connected to a resistive load bank composed of twelve 100-watt lamps arranged in four series groups of three lamps each. This arrangement provides incremental mechanical load control: when no lamps are lit, the motor operates under a no-load (idling) condition, and by progressively lighting each series of lamps, specific load increments (25%, 50%, 75%, and 100%) can be systematically applied. This setup ensures precise, easily adjustable, and repeatable control over the motor's mechanical load, allowing for thorough and consistent data collection across varying operating scenarios and health states.

The induction motor is driven and controlled by a versatile electric drive system capable of operating in different control modes, including direct-on-line (DOL), Scalar Control, and Direct Torque Control (DTC). Scalar Control maintains a constant voltage-to-frequency ratio, making it simple and cost-effective but vulnerable to disturbances such as harmonics and external noise. In contrast, DTC offers superior dynamic performance by directly regulating torque and flux; however, it generates high-frequency switching noise that can mask fault-related patterns. These noise sources present real-world challenges in the dataset, enhancing the relevance of the fault detection process to actual industrial conditions. Motor current signals are continuously recorded at a sampling rate of 10 kHz for 100 seconds under various operating conditions and fault scenarios. The key measuring devices used in this study include a Hall-effect current sensor (LA55-P/SP1), a digital signal processor (DSP) (TMS320F28379D), an oscilloscope, a three-phase voltage and current analyzer, and a PC-based data acquisition system (DAQ). The Hall-effect sensor (LA55-P/SP1) is responsible for measuring the motor's current signal with high precision, ensuring that even small variations in current flow caused by faults are captured. It is installed at the motor's power input and operates based on the Hall effect principle, where the magnetic field generated by the current flow induces a proportional voltage signal, which is then converted into a digital format using the ADC module of the DSP with a resolution of 12 bits.

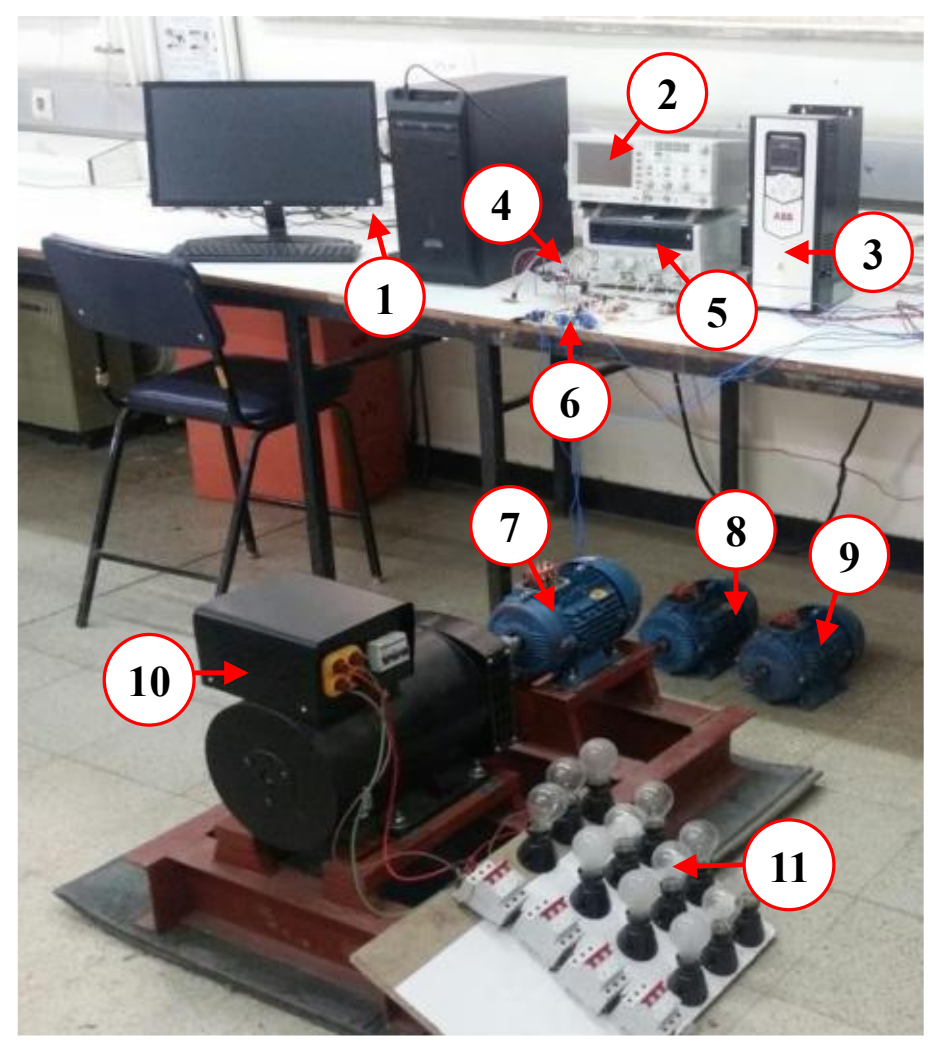

Fig. 2. Experimental fault detection test bench, including: (1) PC, (2) Oscilloscope, (3) Drive, (4) DSP, (5) Power supply, (6) Current sensors, (7) Healthy induction motor, (8) Induction motor with broken rotor bars, (9) Induction motor with eccentricity, (10) Generator, and (11) Electric load.

TABLE I
SPECIFICATION OF THE IM UNDER TEST

| Parameters | Symbol | Values | Units |
|---|---|---|---|
| Normal power | $P_n$ | 1.5 | kW |
| Normal frequency | $f_n$ | 50 | Hz |
| Normal voltage | $V_n$ | 220/380 | V |
| Normal current | $I_n$ | 5.7/3.3 | A |
| Normal speed | $\omega_n$ | 1440 | RPM |
| Normal power factor | PF | 0.81 | - |
| Normal efficiency | $\eta$ | 85.3 | % |
| Number of pole-pairs | p | 2 | - |
| Number of rotor bars | R | 28 | - |
| Air-gap length | $l_g$ | 0.25 | mm |

TABLE II
SPECIFICATION OF BALL-BEARING 6205 NTN

| Parameters | Symbol | Values | Units |
|---|---|---|---|
| Ball diameter | $D_b$ | 7.835 | mm |
| Cage diameter | $D_c$ | 38.5 | mm |
| Number of balls | $N_b$ | 9 | - |
| Contact angle | $\theta$ | 0 | Deg. |

To create an induction motor with a broken rotor bar fault, we disassemble a functioning induction motor and deliberately sever one of the rotor bars using a drilling tool, as illustrated in Fig. 3. This process continues until the number of broken bars reaches five [35].

To replicate the eccentricity fault, we employ two new bearings with modified specifications: reduced outer diameters and increased inner diameters. These alterations are applied to both the drive-end and non-drive-end bearings. We meticulously design three sets of rings, with each set consisting of four distinct rings. The first set is designed to represent normal operating conditions using new bearings. This configuration is depicted in Fig. 4, where four concentric rings are arranged to replicate the optimal condition of the machinery. The remaining two sets each contain four eccentric rings designed to simulate mechanical eccentricity, with each set exhibiting varying degrees of eccentricity. Additionally, a series of concentric rings can be used to simulate dynamic eccentricity [35].

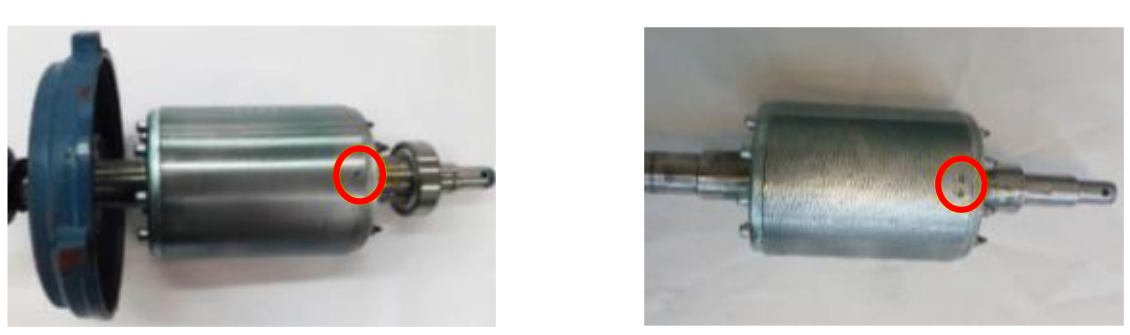

(a) (b)

Fig. 3. Induction motor rotors with broken rotor bars: (a) one broken bar, and (b) two broken bars.

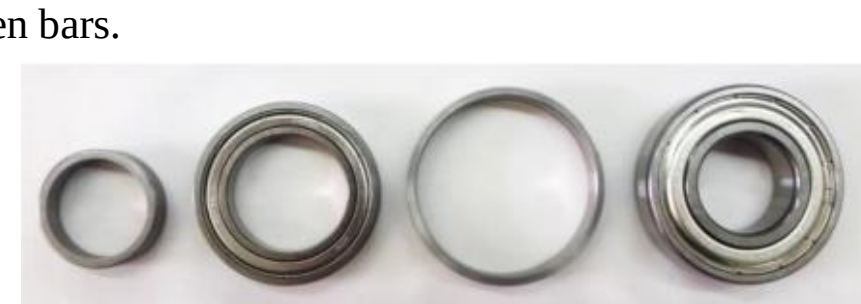

Fig. 4. From right to left: the original ball bearing, one of the outer rings, the new ball bearing, and one of the inner rings.

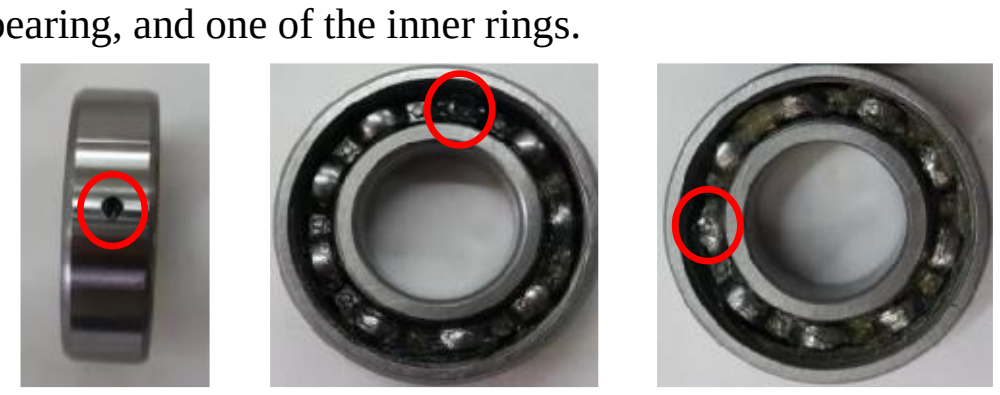

(a) (b) (c)

Fig. 5. Faulty bearings from left to right: (a) outer ring fault, (b) cage fault, and (c) ball fault.

To systematically introduce bearing faults, specific fault types were created using controlled fabrication methods. Outer ring faults were produced by CNC-drilled holes to ensure precision and consistency. Inner ring faults were generated through rotary grinding, simulating the wear and tear commonly encountered in industrial applications. Ball faults were introduced using micro-drilled notches to create localized damage, while cage faults were induced through intentional deformation, including bending and breakage, to replicate real-world failure conditions.

Furthermore, three models of bearing faults are created, as shown in Fig. 5: (a) a hole is made in the outer ring; (b) different cages are either broken or bent due to impact; (c) a scratch is made on the ball using a drill [36]. The process followed for the proposed model is described below.

### *B. Dataset Description*

For designing the matching network architecture, we first select support set consisting of $N$ samples input-output pairs and $M$ samples of query set. After obtaining the support set, it passes through a standard training and application layer, followed by a unidirectional long short-term memory (LSTM) architecture. This process facilitates the learning of the probabilistic label distribution within the support set as we transition between batches. The resulting embeddings are termed fully contextual embeddings. In a recursive manner, the query images undergo a comparable procedure, generating full context embeddings that occupy the same embedding space as those of the support set.

Following the generation of results from both the support set and query set images, these outputs are processed through a kernel and subsequently evaluated using our base learner function. The process of identifying faults across different faulty scenarios is treated as distinct tasks, with each task representing a multiclass classification challenge. The dataset includes 21 current signals and 483 morphological images for bearing faults, 25 signals and 575 images for broken rotor bars, 12 signals and 276 images for eccentricity faults, and 8 signals with 184 images for healthy conditions. Each signal is recorded at 0%, 25%, 50%, 75%, and 100% load levels, generating 23 morphological images per signal. These images serve as essential inputs for training the model, enabling the extraction of key fault-related patterns despite the dataset's limited size.

To enhance generalization, every dataset encompasses various healthy and faulty states and is further segmented into 3,000 subsamples using a subsampling window technique. Each subsample is treated as an independent task and contributes to training the few-shot learning model, ensuring that it effectively learns to classify faults under diverse operating conditions. $T_0$, $T_1$, $T_2$, $T_3$, …, $T_8$ represent different tasks in the dataset, each sourced from the motor running under five different load conditions (0%, 25%, 50%, 75%, and 100%), ensuring variability in operational settings. The corresponding speeds under each load level are detailed in Table III, providing additional insight into how varying mechanical conditions influence the current signals.

Despite the limited dataset, the proposed few-shot learning approach effectively captures critical fault characteristics across multiple load levels, demonstrating high fault detection accuracy even under varying operating conditions. The model's ability to generalize well with a small number of training samples highlights its robustness and adaptability in real-world industrial scenarios where large-scale labeled datasets are often unavailable.

The training of the models based on meta-learning is finalized utilizing the datasets $T_0$, $T_1$, $T_2$, $T_3$ and $T_5$. The details of these tasks can be seen in Table IV. Our proposed model undergoes a training regimen of 500 epochs to attain optimal fault detection capabilities. To enhance the classifier's accuracy, we implement a dynamic learning rate strategy; $\alpha$ is initialized at $10^{-6}$ and gradually escalates to $5 \times 10^{-5}$ throughout the training process. Table V provides a comprehensive list of the model's crucial parameters. Additionally, we set the β to 0.02, which enables the model to swiftly adapt to a variety of previously unseen tasks. The curves illustrating fault detection accuracy under dataset $T_4$ during the 6-way, 1-shot, and 5-shot processes are shown in Fig. 6.

It is evident that after 500 epochs of training, the proposed approach yields impressive results in the 5-shot scenario. The results clearly indicate that the method exhibits excellent performance and high accuracy. Table VI presents a comparison of the proposed method's performance alongside other existing approaches. The suggested model achieves the highest accuracy in the 1-shot, 5-shot, and 10-shot scenarios (98.7%, 99.1%, and 99.4%, respectively), demonstrating strong performance and stability with limited data.

ProtoNet, RelationNet, and MatchNet are meta-learning networks designed for few-shot learning, where labeled data is scarce. ProtoNet classifies samples by computing distances to class prototypes, which aids in identifying subtle differences with limited data. RelationNet, which employs a relation module to measure input-pair similarity, performs comparably well, achieving an accuracy of 98.3% in the 10-shot scenario. MatchNet, although also based on similarity measures, tends to be more variable and less accurate than ProtoNet and RelationNet. The WDCNN-few-shot, adapted from the Wide Deep Convolutional Neural Network, exhibits greater sensitivity to shot variations, resulting in less consistent performance.

In traditional deep learning models trained on the full dataset, ResNet performs best (98.7%) due to its deep architecture and skip connections, which effectively address vanishing gradients. Cap-Net (Capsule Network) follows closely, capturing spatial hierarchies more effectively than CNNs. In contrast, WDCNN focuses on wide-layer feature extraction but achieves lower accuracy (95.8%) in this comparison. ProtoNet and ResNet stand out in the few-shot and full-data categories, respectively, while Siamese-Net and Cap-Net also deliver strong results.

The proposed few-shot model's ability to match or exceed the performance of traditional methods positions it as a strong candidate for fault detection with limited data. However, deep networks like ResNet remain highly effective when abundant labeled data is available, making them ideal for data-rich environments.

TABLE III
RELATIONSHIP BETWEEN MOTOR LOAD AND MOTOR SPEED

| Load (% of full load) | Speed(rpm) |
|---|---|
| 0% | 1492 |
| 25% | 1486 |
| 50% | 1482 |
| 75% | 1473 |
| 100% | 1464 |

TABLE IV
DETAILED DESCRIPTIONS OF EACH TASK

| Task | Number of health state | Sample size | Load | Level of noise (SNR/dB) |
|---|---|---|---|---|
| T0 | 6 | 3000 | 0 to 100% | 0 |
| T1 | 6 | 3000 | 0 to 100% | 0 |
| T2 | 6 | 3000 | 0 to 100% | 0 |
| T3 | 6 | 3000 | 0 to 100% | 0 |
| T4 | 6 | 3000 | 0 to 100% | 0 |
| T5 | 6 | 3000 | 0 to 100% | 0 |
| T6 | 6 | 3000 | 0 to 100% | 2 |
| T7 | 6 | 3000 | 0 to 100% | 4 |
| T8 | 6 | 3000 | 0 to 100% | 6 |

TABLE V
BASELINE CHARACTERISTICS OF THE PARAMETERS OF THE PROPOSED MODEL

| Parameter | Typical Values |
|---|---|
| Batch Size | 32 |
| Number of Inner Steps | 5 |
| Task Distribution | Gaussian |
| Gradient Clipping | 5.62 |
| Support Set Size | 1-5-10 |
| Meta-Objective | CE |
| Inner Optimizer | SGD |
| Outer Optimizer | RMSpromp |

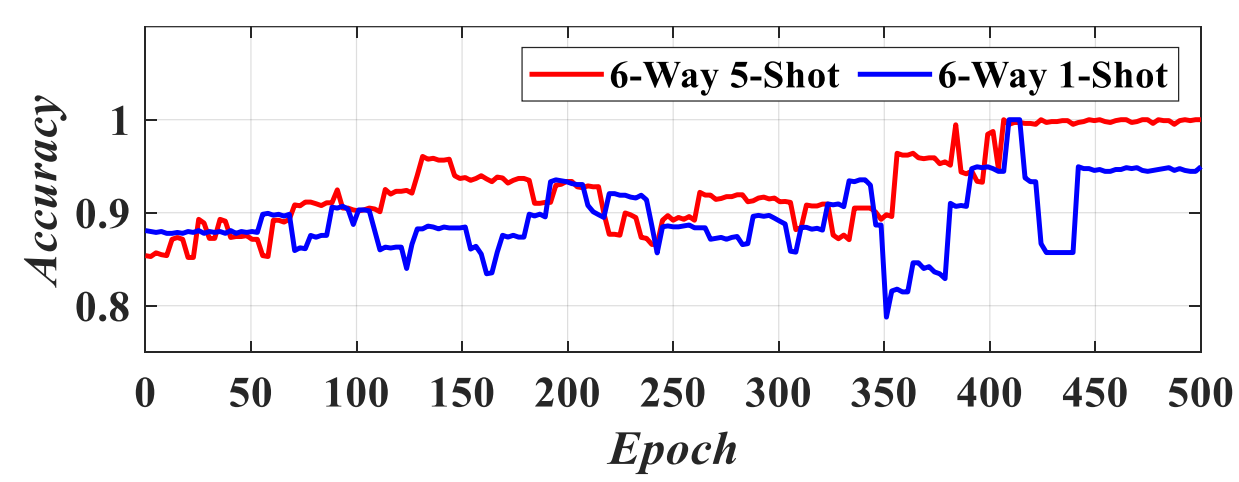

Fig. 6. Visualization of fault detection accuracy during the model training process under 6-way 1-shot and 6-way 5-shot scenarios.

TABLE VI
PERFORMANCE OF VARIOUS FAULT DETECTION METHODS

| Method | Fault detection accuracy (%) | | |
|---|---|---|---|
| Meta-learning base | 1-shot | 5-shot | 10-shot |
| Proposed | 98.7 ± 0.1 | 99.1 ± 0.1 | 99.4 ± 0.0 |
| ProtoNet | 96.3 ± 0.0 | 97.8 ± 0.3 | 98.1 ± 0.3 |
| MatchNet | 94.4 ± 0.4 | 95.4 ± 0.2 | 96.0 ± 0.7 |
| RelationNet | 96.1 ± 0.4 | 98.1 ± 0.0 | 98.3 ± 0.5 |
| WDCNN-few-shot | 97.5 ± 0.6 | 96.6 ± 0.4 | 98.1 ± 0.6 |
| Traditional DL-base | Fault detection accuracy with all dataset (%) | | |
| Siamese-Net | 97.8 ± 0.2 | | |
| Res-Net | 98.7 ± 0.7 | | |
| Cap-Net | 97.9 ± 0.4 | | |
| WDCNN | 95.8 ± 0.3 | | |

### *C. Performance of the Suggested Model Across Different Noise Levels and Unseen Conditions*

In practical scenarios, data collection is influenced by a multitude of factors, including varying operating conditions, human involvement, equipment malfunctions, and numerous other variables. The data gathered from electric machines often contains different levels of noise, which can be characterized as an aggregation of multiple independent random variables, each following distinct probability distributions. Consequently, when an unseen operating condition occurs, the model requires only minimal adjustments to maintain impressive performance, leading to reduced computational costs. However, challenges such as inadequate or incomplete data can arise due to the complexity of the fracture mechanism, the variability of operating conditions, and the difficulties associated with obtaining data from diverse sources and integrating it into a comprehensive database. Therefore, we chose the broken rotor bar as the unseen data, given the limited availability of information resulting from the rarity of broken rotor bars.

We now face an unseen task, and the trained parameters are applied to this challenge. The model's design facilitates rapid convergence, often requiring only a few parameter updates; in some cases, it achieves optimal performance without any modifications at all. The proposed model was tested using dataset $T_4$, achieving accuracies of 88.2 ± 0.2%, 89.4 ± 0.2%, and 90.7 ± 0.4% under 1-shot, 5-shot, and 10-shot conditions, respectively.

A key strength of the proposed model lies in its ability to assess how different noise levels affect fault detection performance, as it relies on real-world noise data. It is well-known that drives generate various harmonics in the current, which are considered noise. In this model, the motor was operated using two drive systems: one employing the SCALAR control technique and the other utilizing direct torque control (DTC) technique, both incorporating noisy data.

To quantify the noise intensity of the data recorded in drive mode, we use a parameter known as the Signal-to-Noise Ratio (SNR), defined as follows [32]:

$$SNR_{dB} = 10log_{10}\left(\frac{P_{signal}}{P_{noise}}\right) = 20log_{10}\left(\frac{A_{signal}}{A_{noise}}\right) \qquad (8)$$

Where $P_{signal}$ and $P_{noise}$ denote the effective power of the signal and noise, respectively, while $A_{signal}$ and $A_{noise}$ represent their average amplitudes.

To ensure that the proposed model demonstrates robust generalization capabilities in the presence of varying noise levels and unfamiliar scenarios, it undergoes training for 800 epochs across multiple tasks. Additionally, the step sizes for adjusting both the main model and the meta-model are set to 0.005 and 0.01, respectively.

To evaluate the proposed method's performance, initial training was conducted on images in their original form without introducing any additional noise. Their performance was subsequently assessed across datasets from multiple tasks ($T_6$ to $T_8$), as shown in Table IV. The meta-learning experiments were carried out using a 6-way 10-shot framework, with performance measured by calculating the average accuracy. The results in Table VII show that our model performs better at 6 dB noise than at 4 dB due to its meta-learning capability, which enables better generalization to unseen conditions. Unlike traditional deep learning models such as WDCNN, which struggle with increasing noise, our approach rapidly adapts to new fault instances while maintaining high accuracy. This reinforces the effectiveness of few-shot learning in real-world industrial settings, where noise variations and unseen data are inevitable. We have included this explanation in the paper to clarify this key advantage. This benefit is further demonstrated in Fig. 7, which illustrates the highest performance levels achieved by leading meta-learning approaches, along with the number of parameter tuning steps required for effective model adjustment to different noise levels. According to Fig. 8, the t-SNE plots indicate that WDCNN, which outperforms other methodologies except for the proposed one, somewhat struggles to distinguish between different fault categories, while the proposed model demonstrates satisfactory proficiency in grouping similar data within each fault category.

The confusion matrices in Fig. 9 and the performance metrics in Table VIII demonstrate the robustness of our MAML-based fault diagnosis model. With F1-scores, precision, and recall exceeding 98% across most fault types, the model ensures reliable classification even under varying noise conditions. Notably, its high adaptability to unseen faults and strong generalization capability outperforms conventional deep learning methods. These results validate the effectiveness of our few-shot learning approach in handling real-world industrial noise, maintaining high accuracy with minimal fine-tuning.

TABLE VII
PERFORMANCE OF VARIOUS METHODS UNDER VARIOUS NOISE LEVELS

| Method | Accuracy at different SNR levels (%) | | | |
|---|---|---|---|---|
| | 0 dB | 2 dB | 4 dB | 6 dB |
| Proposed | **99.1** | **90.9** | **84.3** | **89.9** |
| ProtoNet | **81.1** | **74.8** | **73.2** | **80.1** |
| WDCNN | **86.5** | **79.6** | **84.5** | **83.3** |
| CapNet | 79.4 | 74.2 | 73.0 | 70.0 |
| WDCNN-FS | 76.1 | 73.9 | 71.3 | 68.3 |

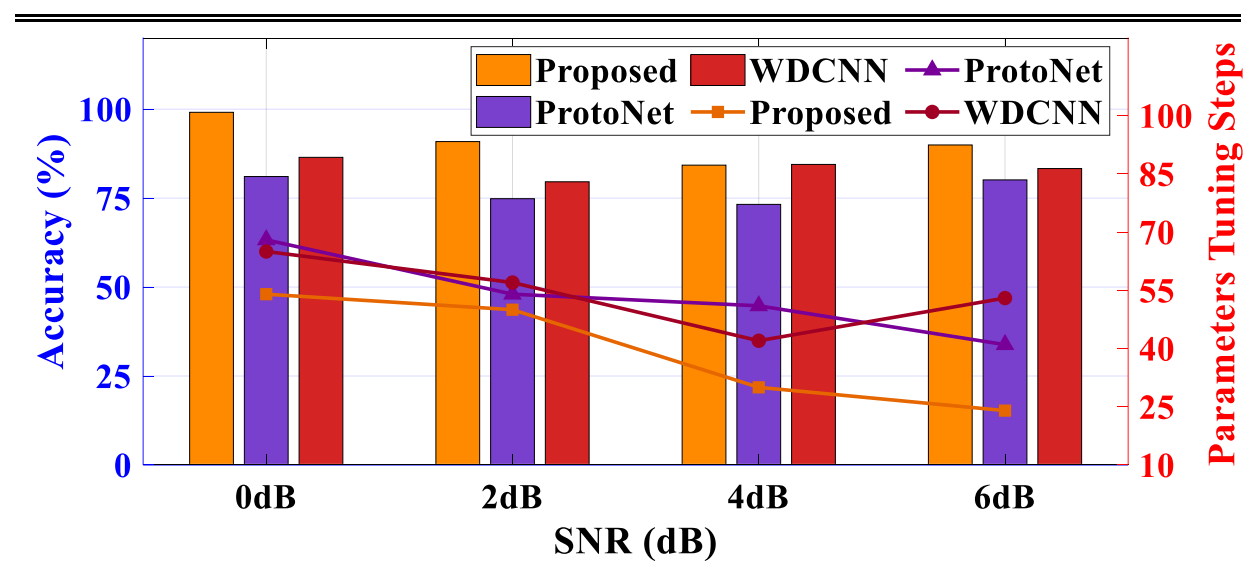

Fig. 7. Performance of different models under different noise levels (6-way 10-shot).

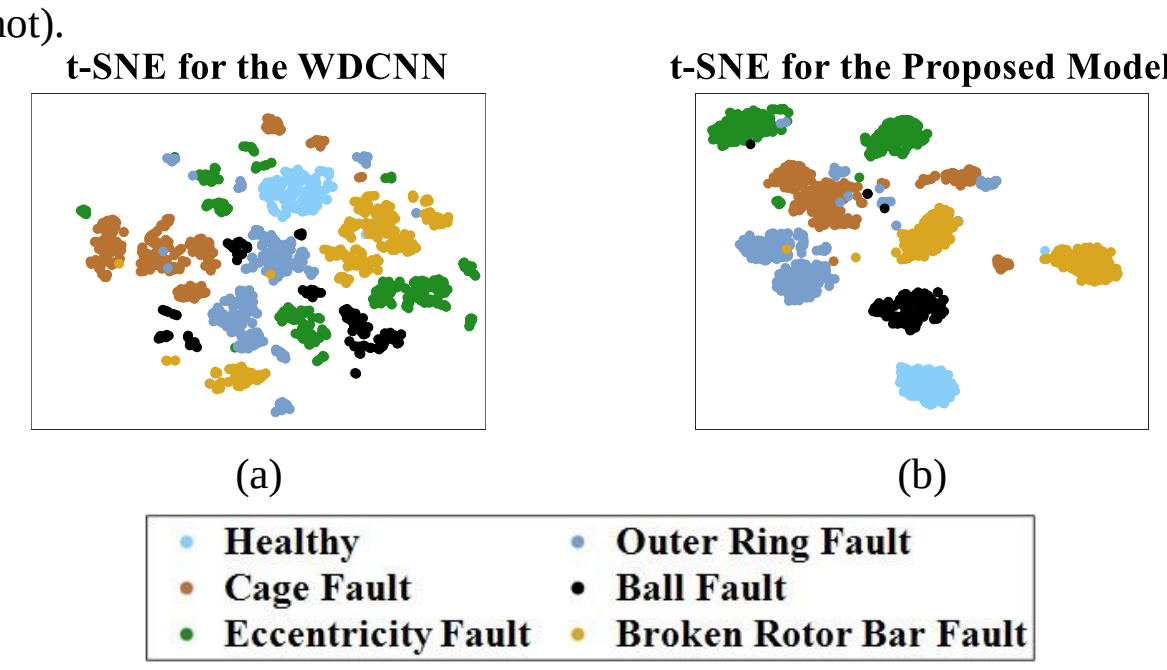

Fig. 8. Feature visualization via t-SNE for WDCNN and the proposed model

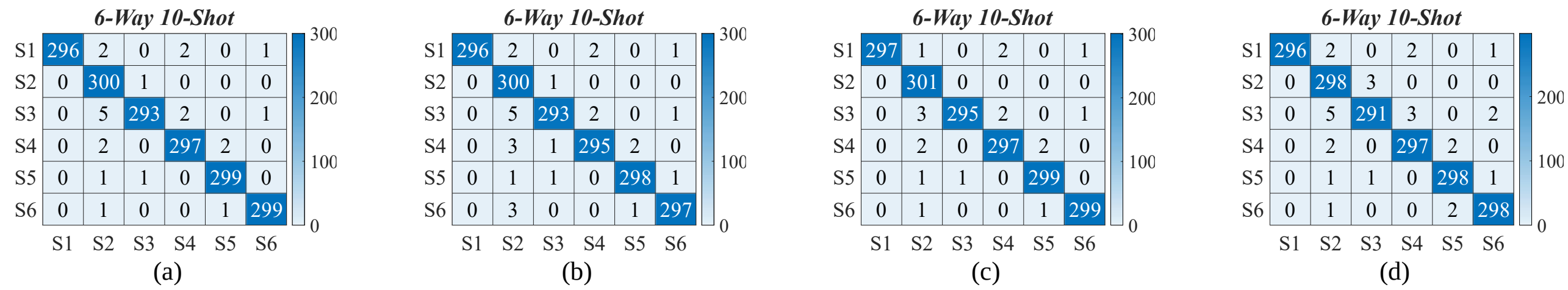

Fig. 9. Confusion matrices of the proposed model for different situations: (a) Combination of Task 4 and Task 8, (b) Combination of Task 4 and Task 6, (c) Task 4, and (d) Combination of Task 4 and Task 7. The designations of the states are as follows: (S1) Ball fault, (S2) Broken rotor bar fault, (S3) Cage fault, (S4) Eccentricity fault, (S5) Healthy condition, and (S6) Outer ring fault.

TABLE VIII
MODEL PERFORMANCE METRICS FOR FAULT CONDITIONS

| | F1-score | Recall | Precision | Accuracy |
|---|---|---|---|---|
| A | 0.987835 | 0.987818 | 0.987997 | 0.987818 |
| B | 0.985088 | 0.98505 | 0.985322 | 0.98505 |
| C | 0.990041 | 0.990033 | 0.990144 | 0.990033 |
| D | 0.984507 | 0.984496 | 0.984633 | 0.984496 |

In summary, the results from this section demonstrate that the proposed model can reliably and accurately identify different health states of induction motors, even with a limited sample size across varying operating conditions. Moreover, the model effectively mitigates the impact of diverse noise levels on fault diagnosis performance, enabling accurate detection of unfamiliar scenarios with minimal adaptation steps.

### *D. Comparison with the State-of-the-Art in Meta-Learning-Based Fault Diagnosis*

In addition, most research conducted at the forefront of knowledge is compared with this method. Jiao Chen et al. [37] proposed a novel approach to addressing the few-shot fault detection problem, which includes a comprehensive framework for detecting bearing faults. Hao Su et al. [38] introduced a fault diagnostic framework called semi-supervised temporal meta-learning (SSTML) for wind turbine bearing fault detection, effectively utilizing both extensive unlabeled data and limited annotated data. Zhiwei Zheng et al. [39] developed a new bearing fault diagnosis method based on meta-learning, tailored for different operating conditions using the AI length and Gram angle field (AI-GAF) data preprocessing technique. Jianjun Chen et al. [20] presented a meta-learning approach for detecting health states in electric machines, designed for rapid adaptation to effectively identify bearing faults.

Compared to the methods described in [20], [37], [38], and [39], our proposed method offers several notable advantages. Firstly, it not only detects bearing faults but also identifies broken rotor bars and eccentricity faults. This capability is significant, as broken rotor bars and eccentricity faults constitute a substantial proportion of mechanical faults in induction motors. Detecting eccentricity faults can be particularly challenging due to their similarity to bearing faults, making this feature a considerable improvement.

When compared to the methods presented in [37] and [38], our proposed model demonstrates significantly higher accuracy, even in scenarios involving noisy or unseen data. This robustness to noise further underscores its effectiveness across varying operating conditions. Additionally, unlike the technique described in [39], which relies on vibration signals, our method utilizes current signals as input. Current signal data have the advantage of being consistently available, and our model eliminates the need for mathematical transformations on the raw input signals, as seen in [39]. This approach simplifies the process and enhances the computational speed of our proposed method.

## IV. CONCLUSION

In conclusion, meta-learning has shown significant promise as an effective approach for fault detection in induction motors. By leveraging machine learning and data-driven techniques, meta-learning algorithms have demonstrated their ability to detect faults efficiently. The future prospects for meta-learning in fault detection are indeed exciting, with anticipated improvements in performance and efficiency as technology continues to advance. The increasing availability of big data and enhancements in computational power will enable meta-learning models to be trained on larger datasets, allowing them to learn and adapt effectively to various fault scenarios.

To address challenges related to small sample sizes and poor generalization in models, this research presents a meta-learning framework focused on diagnosing the health conditions of electric machines, emphasizing rapid adaptation to new data. Experiments conducted using a dedicated dataset demonstrate that the proposed model achieves performance comparable to the latest advancements in the field. It exhibits nearly perfect accuracy in diagnosing health conditions, even with limited sample sizes across diverse operating scenarios. Furthermore, it overcomes the limitations of conventional deep learning techniques, which often suffer from overfitting, and adapts swiftly to new environments, reducing computational demands while showcasing enhanced generalization capabilities.

The model's exceptional fault diagnosis abilities, even in noisy settings, effectively identify the health states of induction motors. Findings indicate that most diagnostic errors arise from incorrect assessments of fault conditions.